\documentclass[5p,twocolumn]{elsarticle}

\usepackage{graphicx}
\usepackage{dcolumn}
\usepackage{pifont}
\usepackage{bm}
\usepackage{multirow}
\usepackage{amsmath}
\usepackage{float}
\usepackage{txfonts}
\usepackage[version=3]{mhchem}

\journal{Elsevier}

\begin{document}
\title{First-principles study on the chemical decomposition of inorganic perovskites \ce{CsPbI3} and \ce{RbPbI3} at finite temperature and pressure}

\author[kimuniv-m,kimuniv-n]{Un-Gi Jong}
\author[kimuniv-m]{Chol-Jun Yu\corref{cor}}
\ead{ryongnam14@yahoo.com}
\author[kimuniv-m]{Yun-Hyok Kye}
\author[kimuniv-n]{Chol-Ho Kim}
\author[kimuniv-s]{Son-Guk Ri}
\author[hongkong]{Yue Chen}

\cortext[cor]{Corresponding author}

\address[kimuniv-m]{Department of Computational Materials Design, Faculty of Materials Science, Kim Il Sung University, \\ Ryongnam-Dong, Taesong District, Pyongyang, Democratic People's Republic of Korea}
\address[kimuniv-n]{Natural Science Centre, Kim Il Sung University, Ryongnam-Dong, Taesong District, Pyongyang, Democratic People's Republic of Korea}
\address[kimuniv-s]{Department of Semiconductor Engineering, Faculty of Materials Science, Kim Il Sung University, \\ Ryongnam-Dong, Taesong District, Pyongyang, Democratic People's Republic of Korea}
\address[hongkong]{Department of Mechanical Engineering, The University of Hong Kong, Pokfulam Road, Hong Kong, SAR, China.}

\begin{abstract}
Inorganic halide perovskite \ce{Cs(Rb)PbI3} has attracted significant research interest in the application of light-absorbing material of perovskite solar cells (PSCs). Although there have been extensive studies on structural and electronic properties of inorganic halide perovskites, the investigation on their thermodynamic stability is lack. Thus, we investigate the effect of substituting Rb for Cs in \ce{CsPbI3} on the chemical decomposition and thermodynamic stability using first-principles thermodynamics. By calculating the formation energies of solid solutions \ce{Cs$_{1-x}$Rb$_x$PbI3} from their ingredients \ce{Cs$_{1-x}$Rb$_x$I} and \ce{PbI2}, we find that the best match between efficiency and stability can be achieved at the Rb content $x\approx$ 0.7. The calculated Helmholtz free energy of solid solutions indicates that \ce{Cs$_{1-x}$Rb$_x$PbI3} has a good thermodynamic stability at room temperature due to a good miscibility of \ce{CsPbI3} and \ce{RbPbI3}. Through lattice-dynamics calculations, we further highlight that \ce{RbPbI3} never stabilize in cubic phase at any temperature and pressure due to the chemical decomposition into its ingredients \ce{RbI} and \ce{PbI2}, while \ce{CsPbI3} can be stabilized in the cubic phase at the temperature range of 0$-$600 K and the pressure range of 0$-$4 GPa. Our work reasonably explains the experimental observations, and paves the way for understanding material stability of the inorganic halide perovskites and designing efficient inorganic halide PSCs.
\end{abstract}


\maketitle

\section{Introduction}
Recently, research interest in inorganic halide perovskites \ce{ABX3} (A $=$ Cs, Rb; B $=$ Pb, Sn; X $=$ I, Br, Cl) has been rapidly growing for applications in photovoltaic solar cells. This is due to their potential high stability against decomposition upon exposure to humidity, light and high temperature~\cite{Eperon,Davis,Lee,Saliba,Duong,Zhang,Saliba2}, as well as favourable properties for light-absorbing materials in perovskite solar cells (PSCs)~\cite{Eperon1,Ahmad,Shum,Chen,Chung,Trots}. Like the organic-inorganic hybrid halide PSCs, moreover, the all-inorganic halide PSCs have a promising advantage of easy fabrication in various ways such as solution processing and vapor deposition techniques~\cite{Liu, Burschka}. These raise hope to make it reality to commercialize PSCs using such inorganic perovskites in the near future.

Through recent theoretical and experimental works, it has been found that phase-stable cubic \ce{CsPbBr3} has too wide band gap to be used as light-absorbing materials for solar cells~\cite{Kulbak,Heidrich}, while cubic $\alpha$-phase \ce{CsPbI3} has an optimal band gap of 1.73 eV for a top cell absorber in tandem solar cells~\cite{Eperon1,Bekenstein}. However, the cubic \ce{CsPbI3} was reported to be phase-unstable, and that is why it can readily transform to the orthorhombic $\delta$-phase, which has a large band gap over 2.7 eV at room temperature~\cite{McMeekin,Moller}. Nevertheless, the cubic $\alpha$-phase \ce{CsPbI3} quantum dots have been turned out to be stable and synthesized by carefully controlling the size of quantum dot. Using these \ce{CsPbI3} quantum dots as a light-absorber, the PSCs have been made, exhibiting a power conversion efficiency (PCE) of up to 10\%~and relatively higher stability~\cite{Swarnkar,Zhang2017}. Based on these experimental findings, it was suggested that nanoscale \ce{CsPbI3} including 2D nanoplates~\cite{Bekenstein,Song} and colloidal nanocrystalline films~\cite{Davis,Protesescu,Gomez,Nedelcu,Akkerman} stabilize the cubic phase. On the other hand, tin-based perovskite \ce{CsSnI3} has been found to possess a lower band gap of 1.3 eV, high optical absorption coefficient of $\sim$10$^4$ cm$^{-1}$ and low exciton binding energy of 18 meV, being smaller than that of the typical organic-inorganic hybrid perovskite \ce{CH3NH3PbI3}~\cite{Shum,Chen,Chung,Song1,Chung12,Borriello}. In addition, \ce{CsSnI3} has an important advantage of non-toxicity because it is a lead-free compound. In spite of these advantages, the PCE of \ce{CsSnI3}-based PSCs was experimentally reported to be too low ($\sim$2\%), possibly due to the unharmonious alignment of band energy level between light-absorber and charge extracting material, and the thermodynamic phase instability~\cite{Chen12,Sabba15,Kumar14}.

There exist several first-principles works for these inorganic perovskites, highlighting the success and failure in their applications to PSCs~\cite{RYang,Jung,Hendon,Brgoch,Murtaza,Silva,Huang,Afsari,Ilyas}. In the previous works carried out by Walsh {\it et al.}~\cite{RYang,Jung}, it was concluded through the lattice-dynamics of \ce{CsSnX3} and \ce{CsPbX3} (X=F, Cl, Br, I) that all the cesium-lead and cesium-tin halide perovskites exhibit a vibrational instability associated with octahedral tilting in the high-temperature cubic phase. This octahedral tilting was found to cause a positive band gap deformation, leading to a larger band gap compared with the cubic perovskite structure. The effect of chemical substitution of Rb for Cs on the structural, thermodynamic, and electronic properties of solid solution Cs$_{1-x}$Rb$_x$SnI$_3$ was also investigated, emphasizing the importance of surface termination in the engineering of the high performance PSCs~\cite{Jung}.

However, the most important question how much the material stability of the all-inorganic perovskites against their chemical decomposition can be enhanced is yet untouched. In fact, this is the most challenging towards the commercialization of PSCs when using the organic-inorganic hybrid halide perovskite absorber, because \ce{CH3NH3PbI3} has been found to decompose into its ingredients \ce{CH3NH3I} and \ce{PbI2} intrinsically as well as extrinsically upon the influence of humidity, temperature and ultraviolet light~\cite{yucj10,yucj12,Buin,Dualeh,Conings,Yang2}. Like organic-inorganic hybrid halide perovskite \ce{CH3NH3PbI3}, the inorganic halide perovskite \ce{CsPbI3} also can be fabricated by spin-coating \ce{CsI} and \ce{PbI2}, and heating to $\sim$600 K~\cite{Eperon}. Therefore, theoretical study on the material stability of \ce{Cs(Rb)PbI3} in opposition to the chemical decomposition into \ce{Cs(Rb)I} and \ce{PbI2} at finite temperature and pressure is of great importance and urgency.

In this work, we first demonstrate the impact of Rb substitution for Cs on thermodynamic stability of mixed rubidium-cesium iodide perovskite \ce{Cs$_{1-x}$Rb$_x$PbI3} by using first-principles thermodynamic calculations. Insights into the thermodynamic stability of \ce{Cs$_{1-x}$Rb$_x$PbI3} are gained by calculating the Helmholtz free energy of mixing \ce{CsPbI3} and \ce{RbPbI3}. We investigate the intrinsic instabilities of these solid solutions by estimating the formation energies of \ce{Cs$_{1-x}$Rb$_x$PbI3} from their components \ce{Cs$_{1-x}$Rb$_x$I} and \ce{PbI2}. We then perform lattice dynamic calculations for \ce{CsPbI3} and \ce{RbPbI3} to obtain their phonon dispersion curves and thus assess their thermodynamic stabilities at finite temperature and pressure. The Gibbs free energy differences between \ce{Cs(Rb)PbI3} and its components \ce{Cs(Rb)I} and \ce{PbI2} are estimated, providing a conclusion that \ce{RbPbI3} is easily decomposed at any temperature and pressure while the decomposition of \ce{CsPbI3} is thermodynamically prohibited at the temperature range from 0 to 600 K and pressure range from 0 to 4 GPa.

\section{\label{method}Computational Methods}
All calculations were carried out using the pseudopotential plane-wave method within the density functional theory (DFT) framework, as implemented in ABINIT package~\cite{abinit09,abinit05} (version 8.4.4). For all the atoms, we used the Troullier-Martins type norm-conserving pseudopotentials (NCPPs)~\cite{Troullier} with valence electronic configurations of Cs-6s$^1$, Rb-5s$^1$, I-5s$^2$5p$^5$, and Pb-5d$^{10}$6s$^2$6p$^2$, constructed by FHI98PP code~\cite{Fuchs}. As increasing the Rb content $x$ from 0.0 to 1.0 with an interval of 0.1, we generated NCPPs of the virtual atoms \ce{Cs$_{1-x}$Rb$_x$} for calculating the decomposition energies and Helmholtz free energies of solid solutions \ce{Cs$_{1-x}$Rb$_x$PbI3} within the virtual crystal approximation (VCA) method~\cite{yucj07,yucj10,yucj12}. The kinetic cutoff energy for plane wave basis set was set to be 40 Ha, and $k$-point meshes of (8$\times$8$\times$8) for structural optimization and (6$\times$6$\times$6) for lattice-dynamics were used. These computational parameters guarantee the total energy convergence of 0.5 meV per unit cell and the phonon energy convergence of 0.1 meV. All the atomic positions were relaxed until the atomic forces were less than 1 meV \AA$^{-1}$. For the exchange-correlation interaction between the valence electrons, the Perdew-Burke-Ernzerhof functional (PBE)~\cite{pbe} within generalized gradient approximation (GGA) was used. In addition, we used the semi-empirical van der Waals (vdW-DF) correction~\cite{vdwDF2}.

Suggesting the chemical reaction for synthesis of the solid solutions \ce{Cs$_{1-x}$Rb$_x$PbI3} as follows,
\begin{equation}
\label{equ_formation}
\ce{Cs$_{1-x}$Rb$_x$I}+\ce{PbI2} \rightarrow \ce{Cs$_{1-x}$Rb$_x$PbI3}
\end{equation}
the intrinsic instability can be estimated as the DFT total energy difference as follows, 
\begin{equation}
\label{eq_form}
E_f=E_\ce{Cs$_{1-x}$Rb$_x$PbI3}-(E_\ce{Cs$_{1-x}$Rb$_x$I}+E_\ce{PbI2})
\end{equation} 
where $E_\ce{Cs$_{1-x}$Rb$_x$PbI3}$, $E_\ce{Cs$_{1-x}$Rb$_x$I}$, and $E_\ce{PbI2}$ are the DFT total energies of \ce{Cs$_{1-x}$Rb$_x$PbI3}, \ce{Cs$_{1-x}$Rb$_x$I}, and \ce{PbI2} compounds, respectively.

To assess the miscibility of \ce{CsPbI3} and \ce{RbPbI3} for composing the solid solutions \ce{Cs$_{1-x}$Rb$_x$PbI3}, the Helmholtz free energy of mixing for each composition is calculated as follows,
\begin{equation}
\label{equ_deltaF}
\Delta F=\Delta U-T\Delta S
\end{equation}
where $\Delta U$ and $\Delta S$ are the internal energy and entropy of mixing, and $T$ is the absolute temperature. The internal energy of mixing is calculated according to the following equation,
\begin{equation}
\label{equ_deltaU}
\Delta U=E_\ce{Cs$_{1-x}$Rb$_x$PbI3}-xE_\ce{RbPbI3}-(1-x)E_\ce{CsPbI3}
\end{equation}
The entropy of mixing can be calculated in the homogeneous limit as follows,
\begin{equation}
\label{equ_deltaS}
\Delta{S}=-k_\text{B}\left[x\ln(x)+(1-x)\ln(1-x)\right]
\end{equation}
where $k_\text{B}$ is the Boltzmann constant. It should be noted that this approach was proven to be successful for both the all-inorganic and organic-inorganic perovskite solid solutions~\cite{Jung,Yi,Brivio16}.

To gain more insightful result for thermodynamic stability, the contribution of lattice vibration should be included through a calculation of phonon dispersion curve and phonon DOS. We applied the density functional perturbation theory (DFPT) method~\cite{baroni} to the phonon calculations of \ce{Cs(Rb)PbI3}, \ce{Cs(Rb)I} and \ce{PbI2}. The calculated phonon DOS was used in estimation of the thermodynamic potential functions such as Helmholtz and Gibbs free energies, which are at finite temperature and pressure for each crystalline compound. Based on these thermodynamic potential functions of \ce{Cs(Rb)PbI3}, \ce{Cs(Rb)I} and \ce{PbI2}, we built the $P-T$ diagram for decomposition of \ce{Cs(Rb)PbI3} into \ce{Cs(Rb)I} and \ce{PbI2}.

The Gibbs free energy $G(T,~P)$ of a compound as a function of temperature $T$ and pressure $P$ is given as follows,
\begin{equation}
\label{equ_G}
G(T,~P)=F(T,~V)+PV
\end{equation}
where $F(T,~V)$ is the Helmholtz free energy as a function of temperature and volume $V$. For the $PV$ term, we calculated the total energies as systematically varying the volume, fitted them into the empirical equation of state (EOS) for solid to obtain $E(V)$ function, and estimated the pressure by conducting a differentiation like $P=-(\partial E/\partial V)_T$~\cite{Yu07}. Within the adiabatic approximation, $F(T,~V)$ can be separated into ionic vibrational and electronic contributions as follows~\cite{Grabowski,Yu07},
\begin{equation}
\label{equ_F}
F(T,~V) = F_\text{vib}(T,~V) + F_\text{el}(T,~V)\simeq F_\text{vib}(T,~V) + E(T=0~\text{K},~V)
\end{equation}
It should be noted that in the electronic Helmholtz free energy, $F_\text{el}(T,~V)=E(T=0~\text{K},~V)-TS_\text{el}$, the $TS_\text{el}$ term is ignored because the electronic temperature effect is negligible for non-metallic systems at the room temperature vicinity~\cite{landau}. Meanwhile, within the quasiharmonic approximation (QHA), the ionic Helmholtz free energy $F_\text{vib}$ can be calculated as follows~\cite{Yu07,Siegel,Grabowski},
\begin{equation}
\label{equ_F_vib}
F_\text{vib}(T,~V)=3Mk_\text{B}T\int_0^{\omega_L}\ln\left\{2\sinh\left[\frac{\hbar\omega(V)}{2k_\text{B}T}\right]\right\}g(\omega)d\omega
\end{equation}
where $\omega(V)$ is the phonon frequency (or phonon energy) as a function of volume, $M$ the atomic mass, $g(\omega)$ the normalized phonon DOS and $\omega_L$ the maximum of the phonon frequencies. Finally, we estimated the thermodynamic stability of \ce{Cs(Rb)PbI3} against the decomposition into \ce{Cs(Rb)I} and \ce{PbI2} by calculating their Gibbs free energy difference as follows,
\begin{equation}
\label{eq_decomG}
\Delta G(T,~P)=G_\ce{Cs(Rb)PbI3}(T,~P)-[G_\ce{Cs(Rb)I}(T,~P)+G_\ce{PbI2}(T,~P)]
\end{equation} 
which is at the finite temperature and pressure.

\section{Results and discussion}
\subsection{Stability and miscibility of solid solutions \ce{Cs$_{1-x}$Rb$_x$PbI3}}
The structural and electronic properties of the inorganic lead-iodide perovskite solid solutions \ce{Cs$_{1-x}$Rb$_x$PbI3} were sufficiently described in the previous studies~\cite{Hendon,Brgoch,Murtaza,Silva,Huang,Afsari,Ilyas}. In particular, we demonstrated that when substituting Rb for Cs in cubic $\alpha-$phase \ce{CsPbI3}, the performance of PSCs using it as a light-absorbing material can be enhanced due to a decrease of band gap and exciton binding energy~\cite{Jong}. Therefore, we tried to tackle the major problem of their material stability in this work.

\begin{figure}[!t]
\begin{center}
\begin{tabular}{@{\hspace{3.5pt}}l}
(a) \\
~~~~~~\includegraphics[clip=true,scale=0.15]{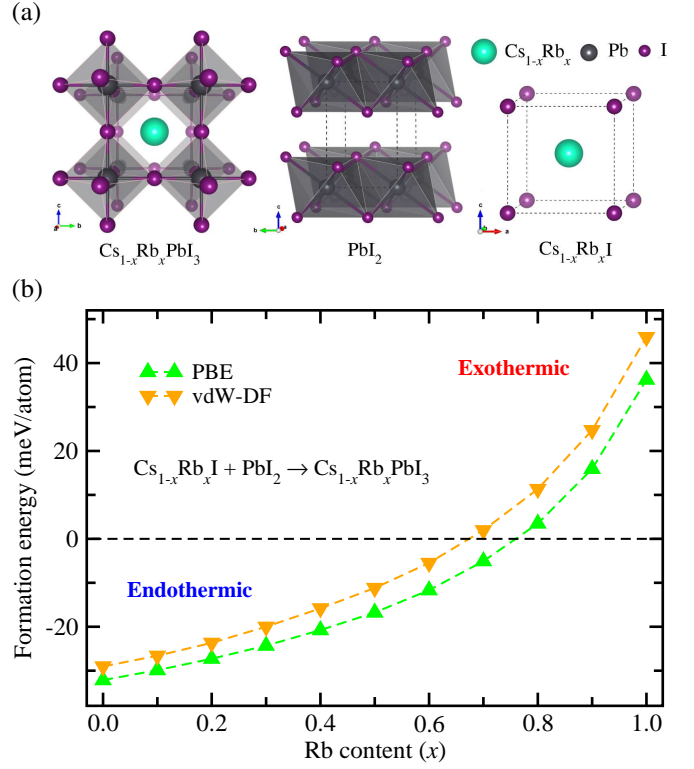} \\
(b) \\
\includegraphics[clip=true,scale=0.54]{fig1b.eps}
\end{tabular}
\caption{\label{fig_Ed}(a) Crystalline structures of cubic \ce{Cs_{$1-x$}Rb_{$x$}PbI3}, hexagonal \ce{PbI2} and cubic \ce{Cs_{$1-x$}Rb_{$x$}I}, and (b) formation energy of \ce{Cs_{$1-x$}Rb_{$x$}PbI3} from their components calculated by using the virtual crystal approximation approach with PBE and vdW-DF functionals.}
\end{center}
\end{figure}

The crystalline structures of \ce{Cs$_{1-x}$Rb$_x$PbI3} and \ce{Cs$_{1-x}$Rb$_x$I} were supposed to be cubic phase with a P\={m}3m space group, and \ce{PbI2} to be hexagonal phase with a P\={3}m1 space group (see Fig.~\ref{fig_Ed}(a)). We carried out the variable cell relaxations of their primitive unit cells. For the cases of solid solutions \ce{Cs$_{1-x}$Rb$_x$PbI3} and \ce{Cs$_{1-x}$Rb$_x$I}, VCA approach was used to treat the primitive unit cell containing one formula unit rather than supercells~\cite{yucj07,yucj10,yucj12}, and the optimized lattice constants were determined as varying the Rb content from $x=$ 0 to 1 with the 0.1 interval, confirming that the Vegard's law is reasonably satisfied and being a good agreement with experiment for the case of \ce{CsPbI3} (6.32 \AA~vs. 6.29 \AA) (see Fig. S1$\dag$).

Then we calculated the formation energies of \ce{Cs$_{1-x}$Rb$_x$PbI3} from their components \ce{Cs$_{1-x}$Rb$_x$I} and \ce{PbI2} using Eq.~\ref{eq_form} with PBE and vdW-DF functionals, as shown in Fig.~\ref{fig_Ed}(b). This formation energy can be regarded as a thermodynamic stability of material at absolute zero of temperature. For the two end compounds, it was found that both the PBE and vdW-DF functionals yield the formation energy to be negative ($\sim$30 meV per atom) for \ce{CsPbI3} while to be positive ($\sim$40 meV) for \ce{RbPbI3}, indicating that \ce{CsPbI3} is intrinsically stable while \ce{RbPbI3} is unstable. As increasing the Rb content, the formation energy increases gradually, and eventually changes from negative to positive value at $x\approx$ 0.7. This indicates that the chemical decomposition of \ce{Cs_{$1-x$}Rb_{$x$}PbI3} is endothermic at $x<0.7$ while exothermic at $x>0.7$, namely, the solid solution may be decomposed spontaneously without any extrinsic factors when the Rb content is over 0.7. It should be noted that once the compound would be decomposed with a production of \ce{PbI2}, the performance of solar cells could be degraded owing to the large band gap of \ce{PbI2} over 2.5 eV. These findings are in reasonable agreement with the experimental findings that the nanoscale \ce{CsPbI3} could stabilize the cubic $\alpha$-phase~\cite{Davis,Protesescu,Gomez,Nedelcu,Akkerman}, but the cubic \ce{RbPbI3} is rarely observed~\cite{Trots,Saliba}. Therefore, it can be said that although the addition of Rb to the cubic \ce{CsPbI3} is expected to enhance the performance of PSCs, the amount of Rb substitution for Cs should be carefully controlled due to a damage of material stability.

The compositional stability of compound from its components can be interpreted by analyzing charge transfer upon its formation, and to this end, we analyzed Hirshfeld charge of each atom in \ce{CsPbI3} and \ce{RbPbI3} (see Table S1$\dag$). It was found that upon the formation of cubic \ce{Cs(Rb)PbI3} crystal, Pb and Cs (Rb) atoms lose 0.295 (0.283) electron and 0.248 (0.228) electron, respectively, while I atom gains $-$0.543 ($-$0.511) electron. When comparing between \ce{CsPbI3} and \ce{RbPbI3}, the amount of charge transfer in the former compound is slightly larger than in the latter compound, implying that as going from $x$ = 1.0 to $x$ = 0.0, the compositional stability becomes better due to enhancement of charge transfer.

In the assessment of material stability of solid solution \ce{Cs_{$1-x$}Rb_{$x$}PbI3}, the miscibility of mixing \ce{CsPbI3} and \ce{RbPbI3} can also play an important role. In the previous work~\cite{Jong}, we estimated the miscibility of the two constituent materials by a bowing parameter, which is a secondary coefficient in quadratic fitting equation of band gaps of \ce{Cs_{$1-x$}Rb_{$x$}PbI3} as a function of the Rb content $x$. In that work, the bowing parameter was estimated to be 0.071 eV, which is smaller by one order than those of the mixed halide perovskites \ce{CH3NH3Pb(I$_{1-x}$Br$_x$)3}~\cite{yucj10} and \ce{CH3NH3Pb(I$_{1-x}$Cl$_x$)3}~\cite{yucj12}. Such smaller bowing parameter indicates a low compositional disorder and thus a good miscibility between \ce{CsPbI3} and \ce{RbPbI3}. However, the bowing parameter is thought to give only a qualitative estimation for miscibility of mixing.

\begin{figure}[!b]
\begin{center}
\includegraphics[clip=true,scale=0.54]{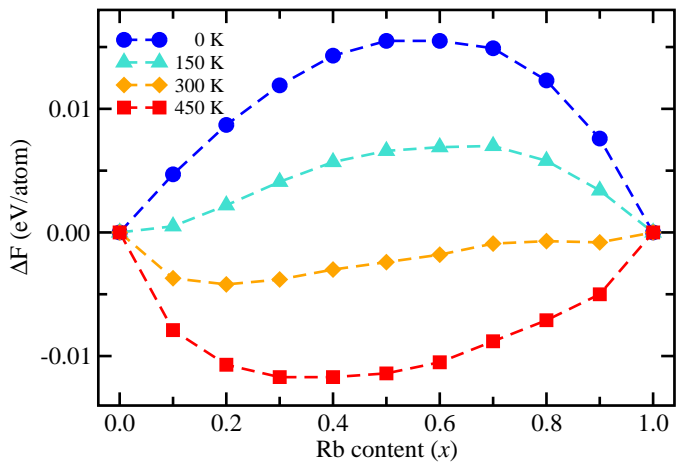} 
\caption{\label{fig_Fm}Helmholtz free energy of mixing \ce{CsPbI3} and \ce{RbPbI3} to form solid solution \ce{Cs_{$1-x$}Rb_{$x$}PbI3} at various temperatures.}
\end{center}
\end{figure}
In order to thermodynamically describe the miscibility between \ce{CsPbI3} and \ce{RbPbI3}, we calculated the Helmholtz free energy of mixing those two materials using Eq.~(\ref{equ_deltaF})-Eq.~(\ref{equ_deltaS}). According to these equations, the free energy differences of the two end compounds \ce{CsPbI3} and \ce{RbPbI3} upon mixing are determined to be zero, and negative free energy of mixing indicates that the solid solution \ce{Cs_{$1-x$}Rb_{$x$}PbI3} is more stable than its individual components while positive values less stable. Figure~\ref{fig_Fm} shows the Helmholtz free energy of mixing as a function of Rb content $x$ at various temperatures. At absolute zero of temperature, the Helmholtz free energies were calculated to be positive in the whole range of Rb content, since the entropy term does not contribute to the free energy. This indicates that it is favourable for the solid solution \ce{Cs_{$1-x$}Rb_{$x$}PbI3} at any Rb content to be separated into its constituents at 0 K. When increasing the temperature, the entropy effect should play an increasing role, leading to negative Helmholtz free energy of mixing, which indicates that solid solutions are gradually stabilized at rising temperatures. In particular, all the solid solutions \ce{Cs_{$1-x$}Rb_{$x$}PbI3} at the whole range of Rb content were predicted to be more stable than the constituents at room temperature of 300 K. It should be noted that vibrational contribution to the free energy is not included since they are expected to give a negligible effect. This result is similar to the case of tin-based iodide perovskites \ce{Cs_{$1-x$}Rb_{$x$}SnI3}~\cite{Jung}. Such a good miscibility for mixing Rb and Cs in these solid solutions can be understood due to a close similarity between Rb and Cs elements. As our calculation reveals a good miscibility for composing \ce{Cs_{$1-x$}Rb_{$x$}PbI3} from the constituents \ce{CsPbI3} and \ce{RbPbI3} at room temperature, it can be thought that the stability of solid solutions is dependent on the stability of the two constituents.

\subsection{Chemical stability of \ce{CsPbI3} and \ce{RbPbI3} at finite temperature and pressure}
We turned our attention to the chemical stability of \ce{CsPbI3} and \ce{RbPbI3} against their decomposition into \ce{CsI}, \ce{RbI} and \ce{PbI2} at finite temperature and pressure. To this end, we carried out a series of SCF calculations as systematically varying unit cell volume for the $PV$ term and DFPT phonon calculations for the $TS$ term, and therefore, estimated the Gibbs free energy of all the compounds, viz. \ce{CsPbI3}, \ce{RbPbI3}, \ce{CsI}, \ce{RbI} and \ce{PbI2}.

\begin{figure}[!b]
\centering
\includegraphics[clip=true,scale=0.55]{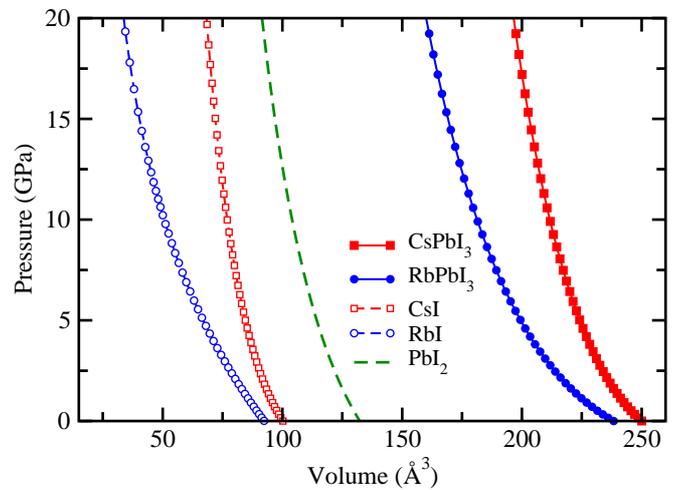}
\caption{\label{fig_PV}Pressure as a function of volume for cubic \ce{CsPbI3}, \ce{RbPbI3}, \ce{CsI}, \ce{RbI} and hexagonal \ce{PbI2}.}
\end{figure}
To estimate the $PV$ term, we calculated the total energies of each compound as increasing the unit cell volume from 0.9$V_0$ to 1.1$V_0$ with 10 volume points, where $V_0$ is a predicted equilibrium volume with the lowest energy. In the case of \ce{PbI2} with the hexagonal lattice, we optimized the lattice constant ratio of $a/c$ with an allowment of atomic relaxation at each fixed volume. Subsequently, we did a fitting of the obtained $E-V$ data into the natural strain 4th-order EOS~\cite{nateos}, determining the equilibrium volume $V_0$ and bulk modulus $B$ (see Fig. S2$\dag$). Then, we derived pressure ranging from 0 to 20 GPa as a function of volume by differentiating the $E(V)$ function as shown in Fig.~\ref{fig_PV}. The lattice constants as a function of pressure are shown in Fig. S3$\dag$. It is natural that the cell volumes and lattice constants for all the compounds decrease as the pressure increases. We note that there might be pressure-induced phase transitions under very high pressure, but they are beyond the scope of this work.

\begin{figure}[!b]
\centering
\includegraphics[clip=true,scale=0.55]{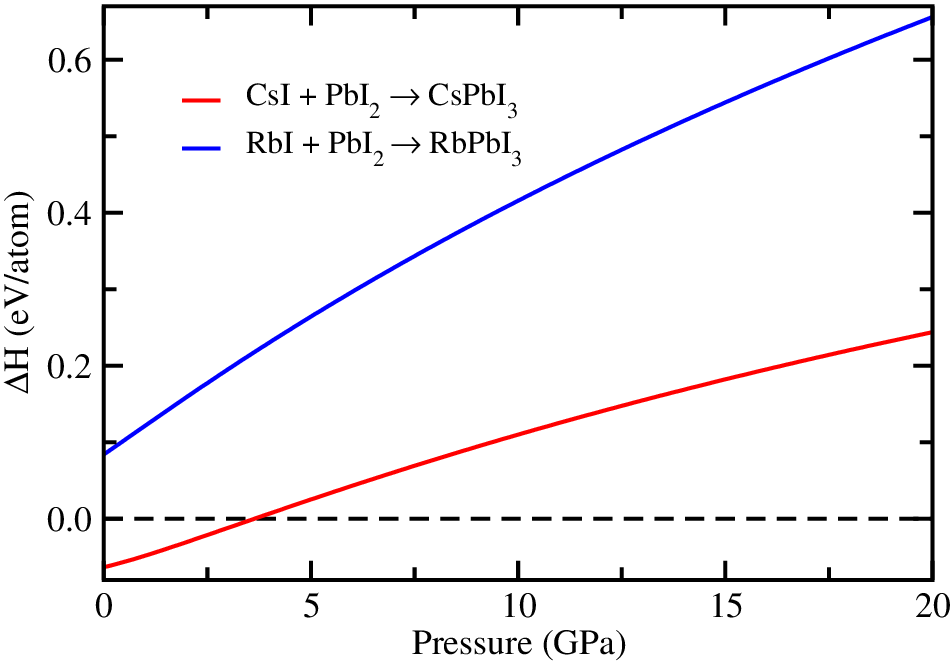}
\caption{\label{fig_P-Hf}Formation enthalpy of \ce{CsPbI3} and \ce{RbPbI3} from their constituents \ce{CsI}, \ce{RbI} and \ce{PbI2} as a function of pressure.}
\end{figure}
Using the calculated $P-V$ data, we estimated the enthalpy $H=E+PV$ for all the compounds. To consider the pure effect of pressure on the chemical decomposition, we calculated the formation enthalpy for the chemical reaction (Eq.~\ref{equ_formation}) as follows,
\begin{equation}
\label{equ_formation_enthalpy}
\Delta H=H_\ce{Cs(Rb)PbI3}-[H_\ce{Cs(Rb)I}+H_\ce{PbI2}]
\end{equation}
As shown in Fig.~\ref{fig_P-Hf}, the formation enthalpies of \ce{CsPbI3} and \ce{RbPbI3} increases gradually as the pressure increases due to the enhancement of the $PV$ term, indicating that their chemical stabilities against the decomposition get worse upon pressing. Moreover, it was found that for the case of \ce{CsPbI3} crystal $\Delta H$ was calculated to be negative at the pressure ranging from 0 to $\sim$4 GPa while for the case of \ce{RbPbI3} to be positive at the whole range of pressure. This indicates that the cubic \ce{CsPbI3} crystal can exist in a stable state at the lower pressure less than $\sim$4 GPa upon the chemical decomposition, but \ce{RbPbI3} can be readily decomposed into its constituents \ce{RbI} and \ce{PbI2} even at the atmospheric pressure. It should be noted that although \ce{CsPbI3} was predicted to be decomposed at the pressure of $\sim$4 GPa, this pressure is 4$\times$10$^4$ times larger than the atmospheric pressure for solar cell operation and thus the pressure has no significant impact on the stability of \ce{CsPbI3} and the performance of PSCs. Then, what about the impact of temperature on the chemical decomposition?

\begin{figure}[!t]
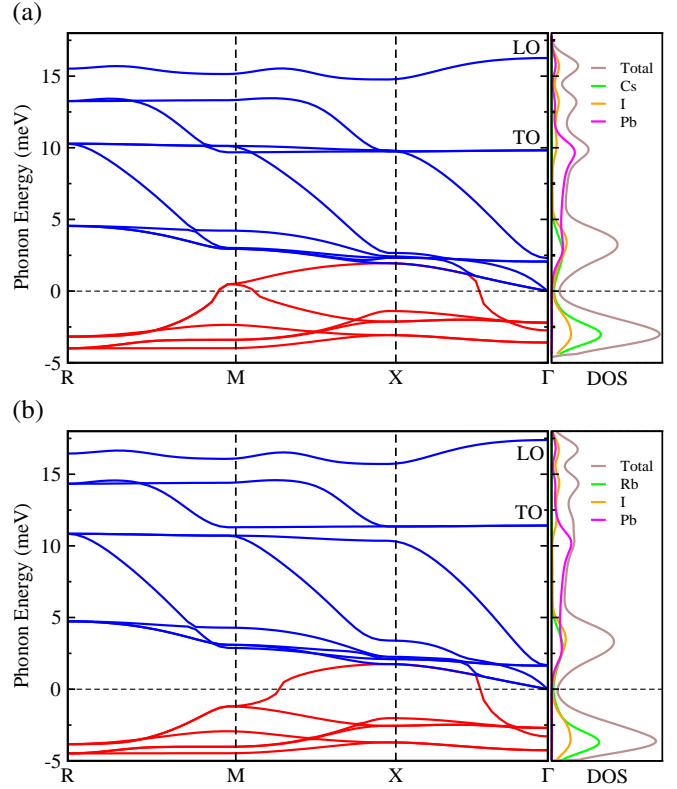

\begin{center}
\begin{tabular}{l}
(a)\\
\includegraphics[clip=true,scale=0.45]{fig5a.eps} \\
(b)\\
\includegraphics[clip=true,scale=0.45]{fig5b.eps} \\
\end{tabular}
\end{center}
\caption{\label{fig_Phonon}Harmonic phonon band structure, and total and atomic resolved phonon density of states for (a) \ce{CsPbI3} and (b) \ce{RbPbI3} in cubic structure, calculated by DFPT method. The phonon dispersion is drawn along the line containing high symmetric points in the vibrational Brillouin zone. Soft phonon modes with imaginary (negative) energy are presented with red colour, and LO-TO splitting appears at $\Gamma$ point.}
\end{figure}

To investigate the finite temperature effect, we calculated the phonon dispersion curves and the corresponding phonon DOS of \ce{CsPbI3}, \ce{RbPbI3}, \ce{CsI}, \ce{PbI} and \ce{PbI2}, which were used in calculation of ionic vibrational contribution to the Helmholtz free energy given in Eq.~\ref{equ_F_vib}. Figure~\ref{fig_Phonon} shows the phonon dispersion curves along the line passing through the high symmetric points (R, M, X, $\Gamma$) in the vibrational Brillouin zone (BZ), and total and atomic resolved phonon DOS in cubic \ce{CsPbI3} and \ce{RbPbI3} (see Fig. S4 for those of \ce{CsI}, \ce{PbI}, and \ce{PbI2}$\dag$). In the calculation of phonon dispersion, we took account the coupling between atomic displacements and the homogeneous electric field existed in the case of polar insulators for longitudinal optic (LO) modes at $\Gamma$ point. Thus, splitting between these modes and the transverse optic (TO) modes (LO-TO splitting) appears at $\Gamma$ point in all the compounds.

For both \ce{CsPbI3} and \ce{RbPbI3} compounds, we observe imaginary phonon modes represented by negative energies in the phonon band structure, so-called {\it soft phonon}, which are a typical characteristics of ferroelectrics in the cubic perovskite structure~\cite{Ghosez,Waghmare,RYang}. On the contrary, all the phonon modes were found to be positive in \ce{CsI}, \ce{RbI} and \ce{PbI2} compounds (see Fig. S4$\dag$).  The identification of the soft phonon indicates a structural instability, which may be associated with the atomic displacement (e.g., in \ce{ATiO3} (A=Ba, Pb), the position of Ti ion is slightly deviated from the centre of an oxygen octahedron~\cite{yucj07}) and/or the octahedral distortion (e.g., in \ce{CsBX3} (B=Pb, Sn; X=F, Cl, Br, I), the halide octahedron is tilted, described as a linear combination of in-plane and out-of-phase rotations along the crystallographic axes~\cite{RYang}). This vibrational or lattice instability is known to drive a phase transition induced by temperature and pressure in the perovskite-type crystals, from the high symmetry cubic phase at high temperature to the lower symmetry phase at lower temperature, including tetragonal, orthorhombic and trigonal or rhombohedral. It is worth noting that for the case of oxide perovskites the lower symmetry phase is a ferroelectric state while the cubic phase represents a paraelectric state. In the case of \ce{CsPbI3}, it was found from experiment~\cite{Trots} that the cubic phase can be stable over room temperature (634 K) while under 298 K the orthorhombic phase is stable.

In Fig.~\ref{fig_Phonon}, both \ce{CsPbI3} and \ce{RbPbI3} compounds are found to exhibit BZ boundary point (R, M, X) instabilities as well as BZ centre ($\Gamma$) point instability. It is worth noting that the BZ boundary distortions are antiferroelectric so that long-range spontaneous polarization is not formed due to a compensation of opposing polarization induced in neighbouring unit cells, whereas $\Gamma$-point instability induces a ferroelectric distortion~\cite{RYang}. From the atomic resolved phonon DOS, it is clear that Cs (Rb) and I atoms are responsible for these lattice instabilities while Pb atom is little related with those. This indicates that the interaction between Cs (Rb) and I atoms can induce the iodine octahedral tilting, of which centre is occupied by Pb atom, and drive the phase transition. This mechanism is contrary to the case of the oxide perovskites like \ce{BaTiO3}, in which the Ti atom placed at the centre of oxygen octahedra is mainly responsible for the lattice instability.

\begin{figure}[!b]
\begin{center}
\includegraphics[clip=true,scale=0.54]{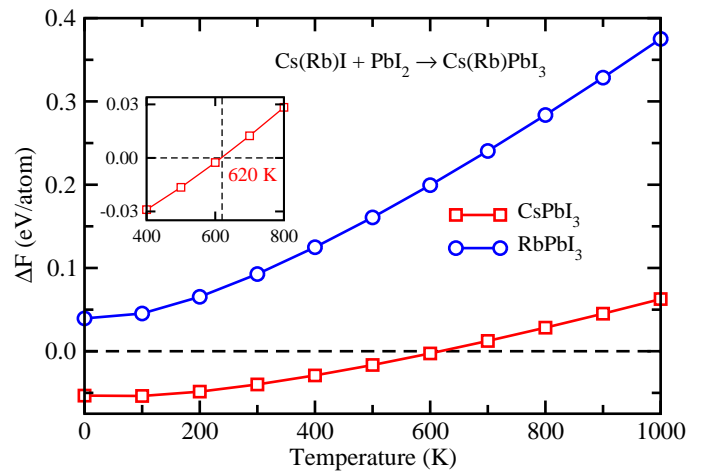}
\caption{\label{fig_deltaF}The Helmholtz free energy difference of \ce{CsPbI3} and \ce{RbPbI3} with respect to \ce{CsI}, \ce{RbI} and \ce{PbI2} as a function of temperature. Inset shows a detail around the sign changing point in the case of \ce{CsPbI3}.}
\end{center}
\end{figure}
It should be emphasised that the lattice instability or thermodynamic instability discussed above is different from the chemical instability for the decomposition of compound. In order to account for the temperature-dependence of chemical decomposition, the Helmholtz free energy difference of \ce{Cs(Rb)PbI3} with respect to \ce{Cs(Rb)I} and \ce{PbI2} was calculated as follows,
\begin{equation}
\label{equ_free_energy_difference}
\Delta F=F_\ce{Cs(Rb)PbI3}-[F_\ce{Cs(Rb)I}+F_\ce{PbI2}]
\end{equation} 
\begin{figure}[!b]
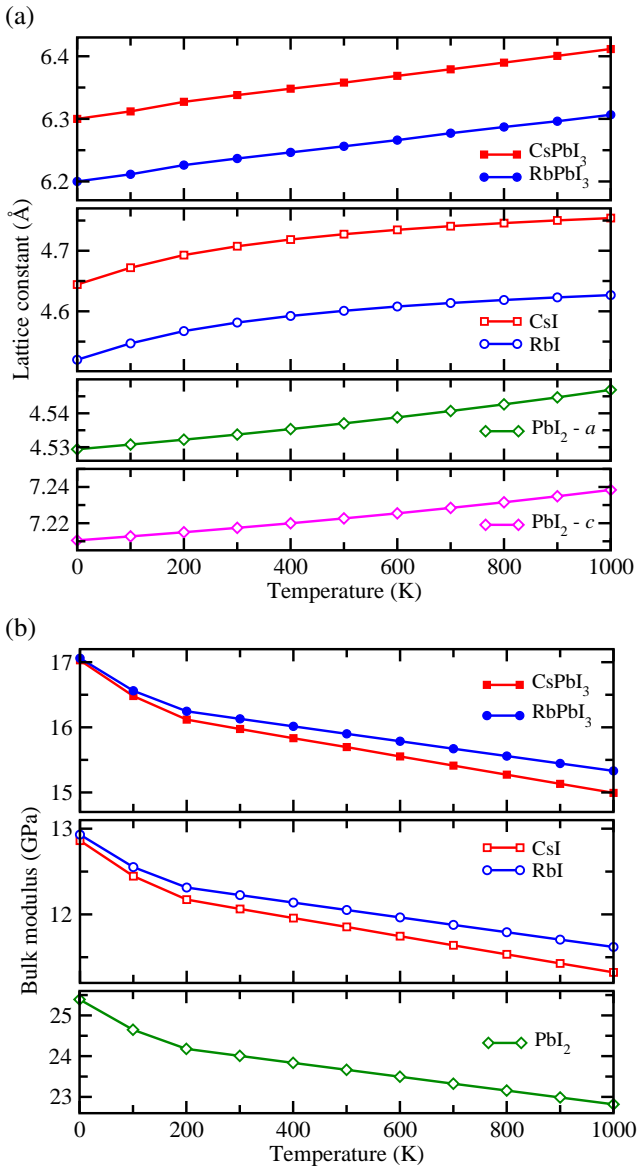

\begin{center}
\begin{tabular}{l}
(a) \\
\includegraphics[clip=true,scale=0.5]{fig7a.eps} \\
(b) \\
~~\includegraphics[clip=true,scale=0.5]{fig7b.eps} \\
\end{tabular}
\end{center}
\caption{\label{fig_latt}(a) Lattice constant and (b) bulk modulus of \ce{CsPbI3}, \ce{RbPbI3}, \ce{CsI}, \ce{RbI} and \ce{PbI2} as the temperature is rising to 1000 K.}
\end{figure}
By performing a post-process of the calculated phonon DOS, we calculated the Helmholtz free energy of all the compounds, that is, \ce{CsPbI3}, \ce{RbPbI3}, \ce{CsI}, \ce{RbI}, and \ce{PbI2}, as increasing the temperature from 0 to 1000 K with an interval of 100 K. Then, the Helmholtz free energy difference $\Delta F$ in \ce{CsPbI3} and \ce{RbPbI3} as a function of temperature was determined and plotted in Fig.~\ref{fig_deltaF}. In Fig.~\ref{fig_latt}, we show the lattice constant and bulk modulus of all the compounds, that is, \ce{CsPbI3}, \ce{RbPbI3}, \ce{CsI}, \ce{RbI} and \ce{PbI2}, as raising the temperature to 1000 K. The bulk modulus was obtained by fitting the $E-V$ data into EOS. When the temperature is rising, the lattice constants are found to increase while the bulk modulus to decrease, being agreed well with the common understanding of thermal expansion of materials.

\begin{figure}[!b]
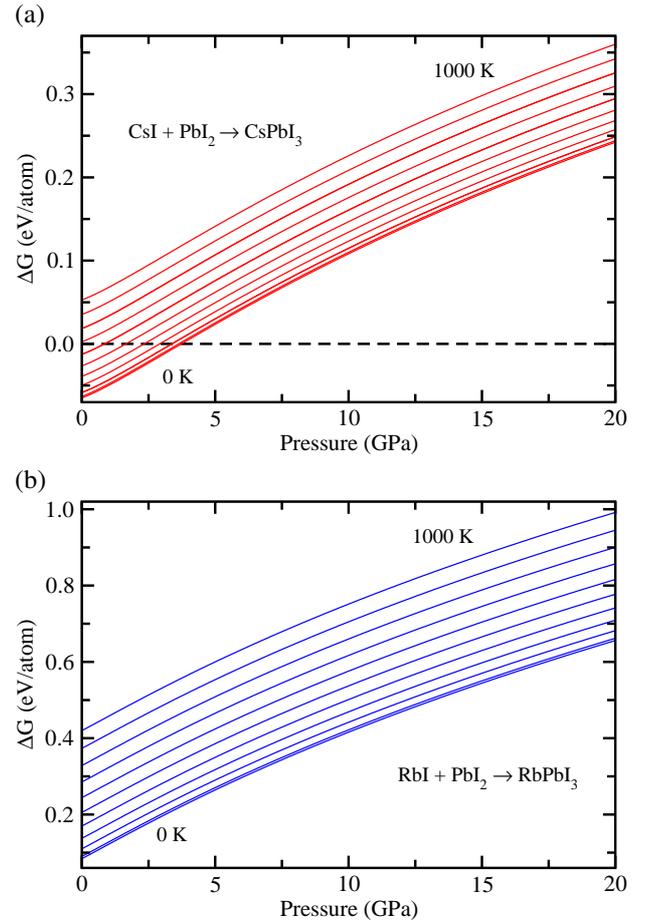

\begin{center}
\begin{tabular}{l}
(a) \\
\includegraphics[clip=true,scale=0.5]{fig8a.eps} \\
(b) \\
\includegraphics[clip=true,scale=0.5]{fig8b.eps} \\
\end{tabular}
\end{center}
\caption{\label{fig_deltaG}Gibbs free energy differences in (a) \ce{CsPbI3} and (b) \ce{RbPbI3} with respect to the \ce{CsI}, \ce{RbI} and \ce{PbI2} as a function of pressure at the temperature range from 0 to 1000 K with an interval of 100 K.}
\end{figure}
According to our calculation, the free energy difference in \ce{RbPbI3} was estimated to be positive at the whole range of temperature, but in the case of \ce{CsPbI3} the sign of energy difference changes from the negative to the positive at $\sim$600 K as it gradually increases when increasing the temperature. This result indicates that \ce{RbPbI3} is unstable upon the chemical decomposition at any temperature, whereas \ce{CsPbI3} is stable upon the chemical decomposition at lower temperature than $\sim$600 K. It should be noted that in the calculation of the free energy the imaginary phonon modes were excluded because they might not be necessarily related to the chemical stability, the aim of this work, though the anharmonic contribution from those should be required to predict the phase transition temperature and pressure. In spite of ignoring the anharmonicity, moreover, our result agrees well with the experimental fact that smooth and uniform thin film of cubic \ce{CsPbI3} can be easily synthesized and stabilized by spin-coating a 1:1 \ce{CsI}:\ce{PbI2} solution and heating to $\sim$600 K~\cite{Eperon}.

To take the temperature and pressure effects together into account, we calculated the Gibbs free energy difference of \ce{CsPbI3} and \ce{RbPbI3} with respect to their constituents \ce{CsI}, \ce{RbI} and \ce{PbI2} at finite temperature and pressure simultaneously. By using the calculated $PV$ and $TS$ term, the Gibbs free energy of all the compounds at the temperature range from 0 to 1000 K and the pressure range from 0 to 20 GPa. Then, the Gibbs free energy difference of each perovskite can be obtained straightforwardly according to Eq.~\ref{eq_decomG}.

In Fig.~\ref{fig_deltaG}, we show the calculated Gibbs free energy difference of \ce{CsPbI3} and \ce{RbPbI3} with respect to their constituents \ce{CsI}, \ce{RbI} and \ce{PbI2} as a function of pressure at the rising temperature from 0 to 1000 K, which can be regarded as $P-T$ diagrams for the chemical stabilities of these materials upon their decompositions. For the case of \ce{CsPbI3}, the obtained $P-T$ diagram contains the individual information for pressure and temperature discussed above. From these $P-T$ diagrams, it is revealed that \ce{RbPbI3} is unstable (that is, it can be readily decomposed) at any temperature and pressure, whereas \ce{CsPbI3} is stable upon the chemical decomposition in the temperature range from 0 to $\sim$600 K and the pressure range from 0 to $\sim$4 GPa. Therefore, we can conclude that when mixing \ce{RbPbI3} with \ce{CsPbI3}, the chemical stability of solid solution \ce{Cs$_{1-x}$Rb$_x$PbI3} could get worse, although the miscibility of mixing them would be good.

\section{Conclusions}
In conclusion, we have investigated the effect of substituting Rb for Cs on the thermodynamic miscibility and chemical stability in the all-inorganic iodide perovskite solid solutions \ce{Cs$_{1-x}$Rb$_x$PbI3} by using the VCA method within the framework of DFT. Through the calculation of formation energy of \ce{Cs$_{1-x}$Rb$_x$PbI3} from its constituents \ce{Cs$_{1-x}$Rb$_x$I} and \ce{PbI2}, it has been found that the Rb content $x\approx$ 0.7 is a turning point at which the decomposition reaction changes from the exothermic to the endothermic, indicating the best performance of PSC at the Rb content $x\approx$ 0.7. Based on the miscibility estimation of mixing the end pure materials \ce{CsPbI3} and \ce{RbPbI3}, we have demonstrated that \ce{Cs$_{1-x}$Rb$_x$PbI3} becomes stable at room temperature due to the increase of configurational entropy. To investigate the chemical stabilities of \ce{CsPbI3} and \ce{RbPbI3} upon their decompositions into \ce{CsI}, \ce{RbI} and \ce{PbI2} at finite temperature and pressure, we have calculated the formation enthalpy, Helmholtz free energy difference, and Gibbs free energy difference of \ce{CsPbI3} and \ce{RbPbI3} with respect to their constituents, resulting in their $P-T$ diagrams for the chemical decomposition. It has been revealed that \ce{RbPbI3} could not be stable in cubic phase at any temperature and pressure due to the ready chemical decomposition into its constituents \ce{RbI} and \ce{PbI2}, whereas \ce{CsPbI3} can be stabilized in cubic phase in the temperature range of 0$-$600 K and the pressure range of 0$-$4 GPa. We believe that this work paves the way for understanding the material stability of the inorganic halide perovskites and thus designing efficient and stable inorganic halide PSCs.

\section*{Acknowledgments}
This work was supported partially by the State Committee of Science and Technology, DPR Korea, under the state project `Design of Innovative Functional Materials for Energy and Environmental Application' (no.2016-20). The calculations have been carried out on the HP Blade System C7000 (HP BL460c) that is owned and managed by the Faculty of Materials Science, Kim Il Sung University.

\section*{Appendix A. Supplementary data}
Supplementary data related to this article can be found at URL.

\section*{\label{note}Notes}
The authors declare no competing financial interest.

\bibliographystyle{elsarticle-num-names}
\bibliography{Reference}

\end{document}